\documentclass{article}

\usepackage{PRIMEarxiv}

\usepackage[utf8]{inputenc} 
\usepackage[T1]{fontenc}    
\usepackage{hyperref}       
\usepackage{url}            
\usepackage{booktabs}       
\usepackage{amsfonts}       
\usepackage{nicefrac}       
\usepackage{microtype}      
\usepackage{lipsum}
\usepackage{fancyhdr}       
\usepackage{graphicx}       
\graphicspath{{media/}}     
\usepackage{subcaption}
\usepackage{tabularx}
\title{ Improved Lung segmentation based on U-Net architecture and Morphological Operations.
}

\author{
  Ali John Naqvi \\
  National University of Sciences and Technology \\
  Islamabad\\
  \texttt{salijohn.ce41ceme@student.nust.edu.pk} \\
   \And
  Abdullah Tauqeer \\
  National University Of Sciences and Technology \\
  Islamabad\\
  \texttt{atauqeer.ee41ceme@student.nust.edu.pk} \\
   \AND
   Rohaib Bhatti \\
    National University Of Sciences and Technology\\
   \texttt{rbhatti.ee41ceme@student.nust.edu.pk} \\
   \And
  Syed Bazil Ali  \\
   National University Of Science and Technology\\
  \texttt{sbazil.ce41ceme@ce.ceme.edu.pk} \\
}

\begin{document}
\maketitle

\begin{abstract}
{An essential stage in computer-aided diagnosis of chest X-rays is automated lung segmentation. Due to rib cages and the unique modalities of each person's lungs, it is essential to construct an effective automated lung segmentation model. This paper presents a reliable model for the segmentation of lungs in chest radiographs.  Our model overcomes the challenges by learning to ignore unimportant areas in the source Chest Radiograph and emphasize important features for lung segmentation. We evaluate our model on public datasets, Montgomery[1] and Shenzhen [2]. The proposed model has a DICE coefficient of 98.3\% which demonstrates the reliability of our model.}
\end{abstract}


\section{Introduction}
\hspace{1.5cm}Medical image segmentation involves extracting the area of interest from medical images. This process involves breaking up an image into informative Regions of Interest (ROIs) that are easy to understand and analyze. Research into medical image segmentation will help processes like pathology monitoring and diagnostic ability improvement by improving accuracy, and precision, and minimizing manual intervention. It is an important step for clinical applications such as Computer-Aided Diagnosis and therapy planning [3]. Image segmentation aims to assign a distinct label to each pixel of an image with the expectation that pixels with the same label would exhibit similar properties. Medical image segmentation provides clinicians a detailed and focused information for accurate diagnosis. The lack of a large, annotated data set [3], the lack of high-quality images, and the extreme variation in images between patients are common problems with medical image segmentation [4]. This demonstrates the need for an efficient, automatic, and general image segmentation model.

\hspace{1.5cm}Computed tomography or magnetic resonance imaging will take over many different types of examinations within the next few years due to the wide range of imaging techniques that are currently available. But with regard to the chest radiograph, this is undoubtedly not the case. The chest can reveal an enormous amount of information about a patient's condition and therefore, chest radiographs are the most popular radiological procedure, making up around 30\% of all radiological treatments. [5]. Lung and heart disorders are studied using chest X-rays, including lung cancer, pneumonia, tuberculosis, and others. Segmentation of chest radiographs is a very important step in computer-aided diagnosis systems. For instance, different changes in the lungs, such as abnormal shape and total area of the lung region, can help with the early diagnosis of life-threatening diseases like cardiomegaly and pleural effusion, among others. High lung segmentation performance is therefore crucial in these applications. However, due to the large diversity in lung images related to age or gender, segmentation is a difficult process.

\hspace{1.5cm}Most of the traditional lung segmentation methods are based on definitive anatomical rules for the segmentation of lung regions from the non-lung region [6]. The majority of these techniques segment lung regions using manually created features. In the era of deep learning, most recent work uses deep learning techniques to segment regions of interest from medical images with higher accuracies [7] [8]. The rapid change in the field of medical picture segmentation is being brought about by deep learning.
In this paper, a deep learning model based on the U-Net architecture is designed for segmenting the lung area. Morphological postprocessing, a group of non-linear operations relating to the appearance or morphology of features in an image, is applied for better extraction of lung region. Through a series of experiments, we show how network architecture and data augmentation may greatly improve the segmentation performance of the model.

\begin{figure}[h]
  \centering
  \begin{subfigure}[b]{0.2\linewidth}
    \includegraphics[width=3cm, height=3cm]{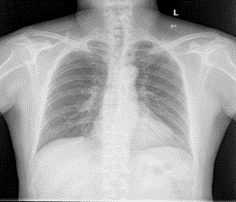}
    \caption{}
  \end{subfigure}
  \begin{subfigure}[b]{0.2\linewidth}
    \includegraphics[width=3cm, height=3cm]{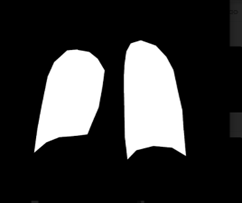}
    \caption{}
  \end{subfigure}
  \caption{Shenzhen data set sample. (a) X-Ray (b) Mask.}
  \label{fig:figure 1}
\end{figure}

\begin{figure}[h]
  \centering
  \begin{subfigure}[b]{0.2\linewidth}
    \includegraphics[width=3cm, height=3cm]{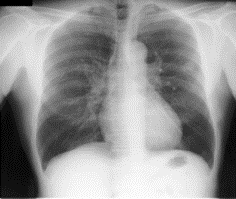}
    \caption{}
  \end{subfigure}
  \begin{subfigure}[b]{0.2\linewidth}
    \includegraphics[width=3cm, height=3cm]{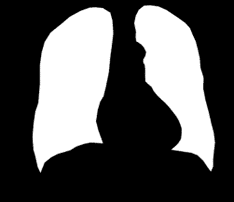}
    \caption{}
  \end{subfigure}
  \caption{Montgomery data set sample. (a) X-Ray (b) Mask.}
  \label{fig:figure 2}
\end{figure}

\section{Method}
\label{Method}
\subsection{Data}
\hspace{1.5cm}Two data sets were used for this experiment. We used the Montgomery data set and the Shenzhen data set. The National Institute of Health and the National Library of Medicine maintain these data sets. In the Montgomery data set, there is a total of 138 images with a size of 4020 * 4892. 58 abnormal images show signs of a wide range of diseases and there are 80 normal cases. Shenzhen X-Ray data set is made up of 662 frontal chest X-Ray with 326 normal cases and 336 abnormal cases. The samples are in PNG format. Images are roughly 3000*3000 in size. Figure 1 and Figure 2 show sample from our data set.

\subsection{Network Architecture}
\hspace*{1.5cm}We have incorporated U-Net architecture into our model. In the U-Net architecture, loss back transmission on higher-level features is replaced by skip connections at the same stage. It ensures that low-level features are combined in the final feature map and permits the mixing of features at various scales. On one side of the architecture, there is an up-sampling process that refines the extracted information. The U-Net architecture is shown in figure 3. In U-Net architecture, repeated convolutions are applied in each layer followed by a max pooling layer which down-samples the image by half in each layer. The max pooling layer picks up the most prominent features in a pool. In the second half of the network, repeated convolutions are followed by up-sampling to match the size of the left side of the network so that they can be concatenated.
U-Net architecture can compute the segmentation of 512 X 512 images in a short time using modern GPUs. Due to its tremendous success, the architecture has undergone many variations and modifications. Some of them include LadderNets [11], attentional U-Nets [12], and recurrent and residual convolutional U-Nets (R2-UNet) [13].
\\
\hspace*{1.5cm}Multiple data augmentation techniques, like zooming, rotating, cropping, etc., are employed on the input data to enhance the sample count of our data. The image is resized to 512 * 512 size after data augmentation. Our approach comprises up-sampling and down-sampling parts, analogous to U-Net architecture. Information from the up-sampling part is combined with the high-resolution low-sampling part to retain the overall segmentation information. After getting results from the U-Net model, the image is fed into a new block where morphological operations and binarization are applied to segmented output to get better results. The entire architecture for our model is shown in Figure 4.
\begin{figure}
  \includegraphics[width=\linewidth]{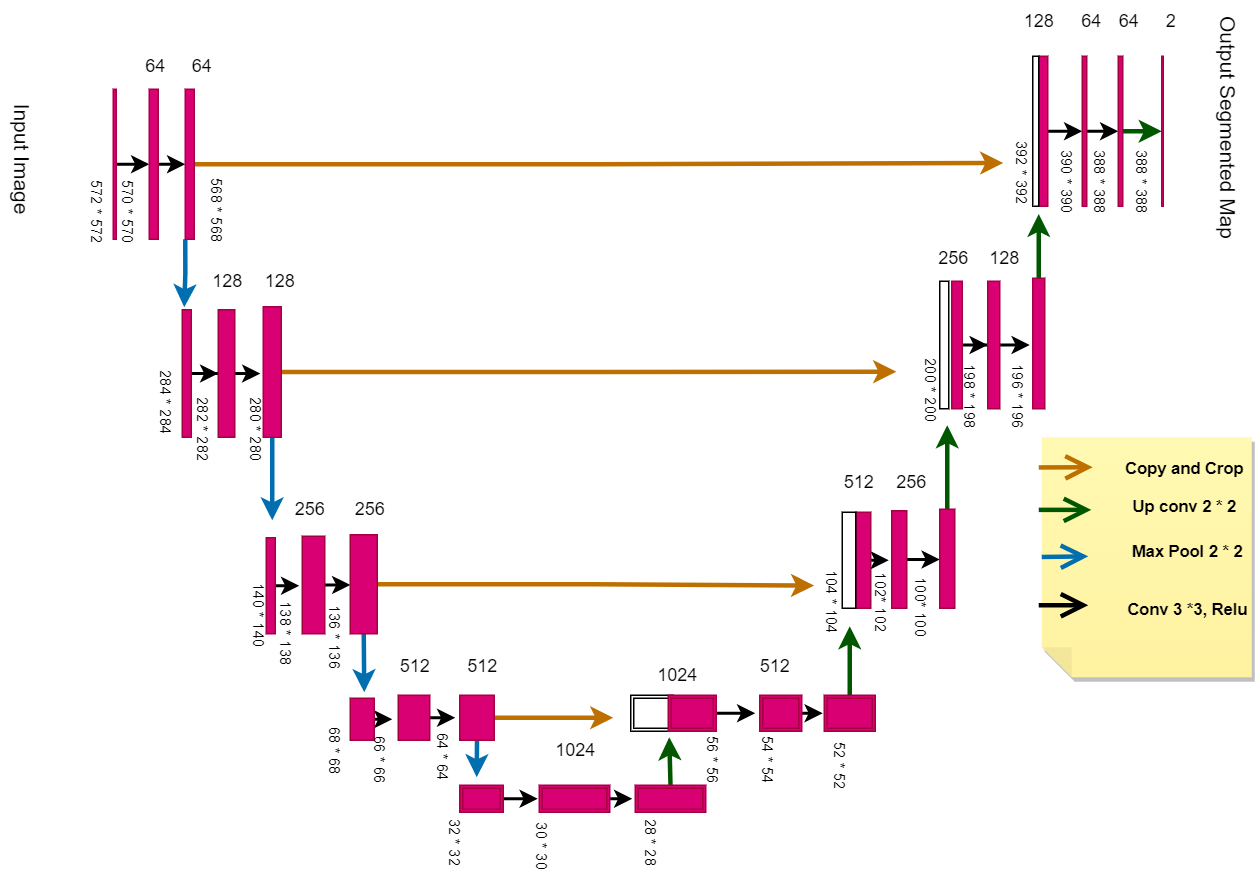}
  \caption{U-Net architecture diagram.}
  \label{fig:architecture}
\end{figure}

\begin{figure}[h]
 \centering
  \includegraphics[width= 10cm, height= 6cm]{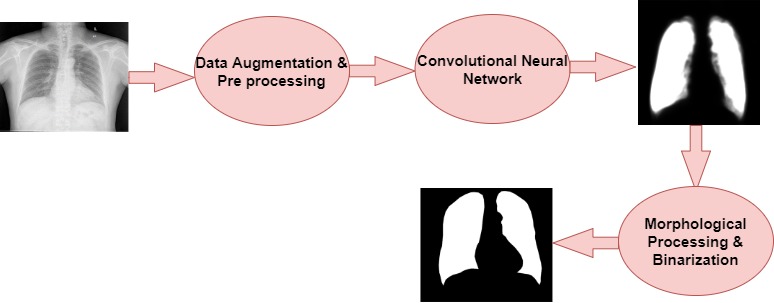}
  \caption{Proposed lung segmentation Model}
  \label{fig:proposed model}
\end{figure}

\section{Experiment}
\hspace{1.5cm}We performed our experiment using Google Colab, a browser application that provides GPU access and allows us to train deep-learning models. We used TensorFlow to develop the proposed model. Each image was resized to 512*512 to match the input of our U-Net model for segmentation. In the Montgomery data set, a separate mask for each lung lobe (right and left lobe) was present and we combine both masks and perform dilation on the combined mask.  We combined both Montgomery and Shenzhen data set samples to make input data set for our model. Overall, 80\% of the data was utilized for training, and 20\% for testing. We trained our model for 50 epochs, batch size of 2, and Adam optimizer. 

\subsection{Results}
\hspace{1.5cm}We report our results in terms of dice coefficients. The dice coefficient is the area where the segmented mask and ground truth overlap. The higher value of dice coefficients indicates a good model for segmentation. As a result, there is little discrepancy between the ground truth and the mask as predicted by the model. The output of our model is seen in Table 1.

\begin{table}[h]
\centering
\caption{Correlation of dice coefficients of our suggested method with the prior state-of-the-art}
\vspace{0.1cm}
\begin{tabular}{|c|c|}
\hline
 Models  & Dice Coefficients \\
 \hline
 ED-CNN [14]& 97.4\\
 \hline
 FCN[15]&97.7\\
 \hline
 Our Model & 98.1\\
 \hline
 
\end{tabular}
\end{table}

\hspace{1.5cm}Our findings imply that data augmentation, pre-processing, and post-processing help to increase our model's dice coefficient by 0.4\%. The segmentation outcomes of our model are shown in Figure 5. The result of our suggested method closely resembles the provided mask.

\begin{figure}[h]
 \centering
  \includegraphics[width= 8cm, height=8cm]{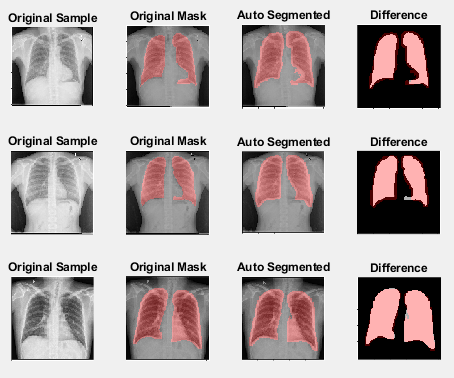}
  \caption{Results of our model in comparison with the ground truth mask.}
  \label{fig:comparison}
\end{figure}

\section{Conclusion}
\hspace{1.5cm}We proposed an architecture based on U-Net which improves the dice coefficient by 0.4\%. We used two publicly available data sets Montgomery and Shenzhen for our experiment. In the Montgomery data set, a mask for each lobe of the lung was present. We combined both lobes of the lung mask into one mask image and perform dilation on the masked image. We add both Montgomery and Shenzhen and used combined data as input data sets. 

\hspace{1.5cm}To achieve excellent results, the work combines a variety of techniques, including data augmentation and morphological operations. We believe that this research significantly advances the field of science since the proposed architecture shows great robustness despite large variations in the input data set. 

\hspace{1.5cm}Furthermore, the inventive ways in which techniques were applied and modified to execute lung segmentation may serve as a model for future works by other authors. It is important to keep in mind that the majority of computer-aided detection systems rely on a trustworthy segmentation technique. Therefore, we consider the recommended approach to be highly relevant and effective.
\subsection{Future Work}
\hspace{1.5cm}Although the suggested method has produced positive results, there are some areas where it could be improved. Here are some existing constraints to our work that could be removed later.

\hspace{1.5cm}The sample size of 800 examinations is considered inadequate for the training of deep learning models, despite the fact that our dataset has an incredible amount of complexity and variety. We propose using additional public or private CXR databases in the trials as future work to support our results. This will offer a more trustworthy lung segmentation approach and provide a better analogy with the existing research.

\hspace{1.5cm}To perform weakly supervised segmentation and to improve the segmentation results we can incorporate a vision transformer [16] into our work to show how transformers perform in this domain.

\bibliographystyle{unsrt}  
\bibliography{references}  

[1] Candemir S, Jaeger S, Musco J, Xue Z, Karargyris A, Antani SK, Thoma GR, Palaniappan K. Lung segmentation in chest radiographs using anatomical atlases with nonrigid registration. IEEE Trans Med Imaging. 2014 Feb;33(2):577-90. doi: 10.1109/TMI.2013.2290491. PMID: 24239990.

[1] Jaeger S, Karargyris A, Candemir S, Folio L, Siegelman J, Callaghan FM, Xue Z, Palaniappan K, Singh RK, Antani SK. Automatic tuberculosis screening using chest radiographs. IEEE Trans Med Imaging. 2014 Feb;33(2):233-45. doi: 10.1109/TMI.2013.2284099. PMID: 24108713.

[2] Jaeger S, Candemir S, Antani S, Wáng YX, Lu PX, Thoma G. Two public chest X-ray datasets for computer-aided screening of pulmonary diseases. Quant Imaging Med Surg. 2014 Dec;4(6):475-7. DOI: 10.3978/j.issn.2223-4292.2014.11.20. PMID: 25525580; PMCID: PMC4256233.

[3] D. Jha, P. H. Smedsrud, M. A. Riegler, P. Halvorsen, T. de Lange,
D. Johansen, and H. D. Johansen, “Kvasir-seg: A segmented polyp
dataset,” in International Conference on Multimedia Modeling (MMM),
2020, pp. 451–462.

[4] F. Zhao and X. Xie, “An overview of interactive medical image segmentation,”
Annals of the BMVA, vol. 2013, no. 7, pp. 1–22, 2013.

[5] B. Van Ginneken, B. M. T. H. Romeny, and M. A. Viergever, "Computer-aided diagnosis in chest
radiography: a survey," IEEE Trans. Med. Imaging, vol. 20, no. 12, pp. 1228-1241, 2001

[6] J. Duryea and J. M. Boone, "A fully automated algorithm for the segmentation of lung fields on
digital chest radiographic images," Med. Phys., vol. 22, no. 2, pp. 183–191, 1995.

[7] Mina Jafari1, Dorothee Auer2, Susan Francis3, Jonathan Garibaldi1, Xin Chen1 “DRU-NET: AN EFFICIENT DEEP CONVOLUTIONAL NEURAL NETWORK FOR MEDICAL IMAGE SEGMENTATION”.

[8] J. Islam and Y. Zhang, "Brain MRI analysis for Alzheimer’s disease diagnosis using an ensemble
system of deep convolutional neural networks". Brain informatics, 5(2), p.2, 2018

[9] O. Ronneberger, P. Fischer, T. Brox, "U-net: Convolutional networks for biomedical image
segmentation". In International Conference on Medical image computing and computer-assisted
intervention, Springer, Cham, pp. 234-241, 2015.

[10] Ronneberger, O., Fischer, P., Brox, T. (2015). U-Net: Convolutional Networks for Biomedical Image Segmentation. In: Navab, N., Hornegger, J., Wells, W., Frangi, A. (eds) Medical Image Computing and Computer-Assisted Intervention – MICCAI 2015. MICCAI 2015. Lecture Notes in Computer Science(), vol 9351. Springer, Cham. ${https://doi.org/10.1007/978-3-319-24574-4_28}$

[11] Mirza, Behroz, Syed, Tahir , Memon, Jamshed , Malik, Yameen. (2017). Ladder Networks: Learning under Massive Label Deficit. International Journal of Advanced Computer Science and Applications. 8. 10.14569/IJACSA.2017.080769.

[12] Oktay, Ozan , Schlemper, Jo , Folgoc, Loic , Lee, Matthew , Heinrich, Mattias , Misawa, Kazunari , Mori, Kensaku , McDonagh, Steven , Hammerla, Nils , Kainz, Bernhard , Glocker, Ben , Rueckert, Daniel. (2018). Attention U-Net: Learning Where to Look for the Pancreas.

[13] Alom, Md. Zahangir , Hasan, Mahmudul , Yakopcic, Chris , Taha, Tarek , Asari, Vijayan. (2018). Recurrent Residual Convolutional Neural Network based on U-Net (R2U-Net) for Medical Image Segmentation.

[14] A. Kalinovsky and V. Kovalev, "Lung image segmentation using deep learning methods and
convolutional neural networks", In XIII International Conference on Pattern Recognition and
Information Processing, 2016.

[15] R. Rashid, M. U. Akram, T. Hassan, "Fully Convolutional Neural Network for Lungs Segmentation
from Chest X-Rays". In International Conference Image Analysis and Recognition, Springer,
Cham, pp. 71-80, 2018.

[16] Dosovitskiy, Alexey , Beyer, Lucas , Kolesnikov, Alexander , Weissenborn, Dirk , Zhai, Xiaohua , Unterthiner, Thomas , Dehghani, Mostafa , Minderer, Matthias , Heigold, Georg , Gelly, Sylvain , Uszkoreit, Jakob , Houlsby, Neil. (2020). An Image is Worth 16x16 Words: Transformers for Image Recognition at Scale.

\end{document}